\documentclass{emulateapj}
\usepackage{graphicx}

\shorttitle{A coronal seismological study}

\shortauthors{Chen et al.}

\usepackage{color}

\begin{document}

\title{A coronal seismological study with streamer waves}

\author{Y. Chen\altaffilmark{1}, S. W. Feng\altaffilmark{1},
B. Li\altaffilmark{1}, H. Q. Song\altaffilmark{1}, L. D.
Xia\altaffilmark{1}, X. L. Kong\altaffilmark{1}, {\&} Xing
Li\altaffilmark{2}} \altaffiltext{1}{Shandong Provincial Key
Laboratory of Optical Astronomy and Solar-Terrestrial Environment,
School of Space Science and Physics, Shandong University at
Weihai, Weihai, China 264209; yaochen@sdu.edu.cn}
\altaffiltext{2}{Department of Physics, University of Wales,
Aberystwyth, SY23 3BZ, UK}

\begin{abstract}
We present a novel method to evaluate the Alfv\'en speed and the
magnetic field strength along the streamer plasma sheet in the
outer corona. The method is based on recent observations of
streamer waves, which are regarded as the fast kink body mode
carried by the plasma sheet structure and generated upon the
impact of a fast CME (coronal mass ejection) on a nearby streamer.
The mode propagates outwards with a phase speed consisting of two
components. One is the phase speed of the mode in the plasma rest
frame, the other is the speed of the solar wind streaming along
the plasma sheet. The former can be well represented by the
Alfv\'en speed outside the plasma sheet, according to a linear
wave dispersion analysis with a simplified slab model of
magnetized plasmas. The radial profiles of the Alfv\'en speed can
be deduced with constraints put on the speed of the solar wind,
which is done by making use of the measurements of streamer blobs
flowing passively in the wind. The radial profiles of the strength
of the coronal magnetic field can be depicted once the electron
density distribution is specified, this is done by inverting the
observed polarized brightness data. Comparing the diagnostic
results corresponding to the first wave trough and the following
crest, we find that both the Alfv\'en speed and magnetic field
strength at a fixed distance decline with time. This is suggestive
of the recovering process of the CME-disturbed corona.
\end{abstract}


\section{Introduction}
The magnetic field plays an important role in physical processes
occurring at all relevant coronal heights. From the photosphere to
the inner corona, the strength of the magnetic field can be
measured by the conventional Zeeman splitting technique. However,
in the outer corona beyond, say, 1.2-1.5 $R_\odot$, the field gets
too weak to be measured directly. Indirect methods available at
present are mostly based on numerical extrapolations or various
types of radio techniques.

With the extrapolation method, the coronal magnetic field
distributions are resolved numerically by extrapolating the
measured photospheric magnetic field, making use of potential
field (Schatten et al. 1969; Schrijver {\&} Derosa 2003), linear
or nonlinear force free field assumptions (e.g., Yan {\&} Sakurai
2000; Wiegelmann, 2008; He {\&} Wang 2008), or solving the
full-set of magnetohydrodynamic (MHD) equations (e.g., Linker et
al. 1999). There also exist various types of techniques employing
radio emissions to derive the coronal magnetic field strength. The
first one utilizes the well-known Faraday-rotation effect acting
on a linearly polarized radio signal passing through coronal
structures. Signals from both extragalactic radio sources (Sakurai
{\&} Spangler 1994; Mancuso {\&} Spangler 1999, 2000; Spangler
2005; Ingleby et al. 2007) and spacecraft radio emitters
(P\"{a}tzold et al. 1987) have been analyzed over the past
decades. To use this method one needs to determine independently
the coronal electron density distribution and the geometry of the
magnetic field. The method applies to the heliocentric distance
range of $3 - 10$ $R_\odot$. The second one is based on the
band-splitting phenomenon observed during Type-II radio bursts
related to shocks driven by coronal mass ejections (CMEs) (Smerd
et al. 1974, 1975; Vrsnak, et al. 2002; Cho et al. 2007). The
phenomenon is interpreted as plasma emissions from downstream and
upstream of the shock front at different frequencies. To implement
this method one needs to apply the MHD shock theory, and presume
the value of the plasma $\beta$, the shock geometry, as well as
the coronal electron density distribution. The method works for
the heliocentric distance range of $1.5 - 3$ $R_\odot$. The third
method makes use of observations of circularly polarized thermal
radio emissions (Sastry 2009), which was recently explored to
estimate the magnetic field strength in coronal streamers at
heliocentric distances of 1.5 and 1.7 $R_\odot$ by Ramesh et al.
(2010). In this paper, we present a novel seismological method to
evaluate the strength of the magnetic field in the outer corona.

Coronal seismology is a way to diagnose the physical parameters of
the corona with observational and MHD theoretical analyses of
waves and oscillations. Here we present a seismological study to
derive the Alfv\'en speed and magnetic field strength with the use
of the so-called streamer waves, reported recently by Chen et al.
(2010) (referred to as paper I hereinafter). The waves were
observed by the Large Angle and Spectrometric Coronagraph (LASCO),
and generated as the aftermath of the CME-streamer interaction
event dated on 6 July 2004. The wave properties like the
wavelength, period, and phase speed, as well as the possibility of
deriving the magnetic field strength with this wave, are already
presented in paper I. Here we briefly discuss the theoretical
basis and procedures of implementing the concerned seismological
study.

The waves are regarded as the fast kink body mode carried by the
plasma sheet structure of a streamer. The phase speed of the mode
have two contributions. One is the phase speed of the mode in the
plasma rest frame, the other is the speed of the solar wind
streaming along the plasma sheet. The former can be well
approximated by the exterior Alfv\'en speed according to a linear
wave dispersion analysis, which will be carried out in the
Appendix of this paper. As a result, the radial profile of the
Alfv\'en speed and that of the magnetic field strength can be
deduced given the speed and density of the solar wind plasmas. In
paper I, by taking these solar wind parameters from a simplified
two-dimensional MHD model accounting for streamers, current-plasma
sheets, and slow-fast winds developed by Chen {\&} Hu (2001), the
magnetic field strengths at 5 and 10 $R_\odot$ are evaluated. It
should be pointed out that the seismological study in paper I is
rather preliminary and incomplete. The purpose of this paper is to
further improve and complete the study.

It will be improved from the following three aspects. First, in
paper I the Alfv\'en speed in the region surrounding the plasma
sheet is set to be equal to the phase speed of the kink mode in
the plasma rest frame. In this paper, we will conduct a parameter
study on dispersion relation of the fast kink body mode with a
simplified slab model of magnetized plasmas given by Edwin {\&}
Roberts (1982), making use of different sets of coronal parameters
prescribed according to available observational diagnoses and
theoretical modelings. From the study, we deduce the appropriate
connection of the kink mode phase speed to the exterior Alfv\'en
speed. Second, we employ available observational results to put
constraints on the flow speed and number density of the solar wind
along the plasma sheet. The former will be constrained by
measurements of plasma blobs, which are structures released
intermittently through streamer cusps and flowing outwards
passively in the wind along the plasma sheet. The feasibility of
using blobs to yield the wind speed has been discussed in several
papers (Sheeley et al. 1997; Wang et al. 1998, 2000; Song et al.
2009; and Chen et al. 2009), and will not be repeated here. The
number density will be derived by inversion of the pB (polarized
brightness) measurements given by LASCO at the time prior to the
CME-streamer interaction event. Third, in paper I only the
magnetic field strengths at two distances were estimated, while in
this paper two sets of radial profiles of both the Alfv\'en speed
and magnetic field strength in the heliocentric range of 3 to 10
$R_\odot$, corresponding to two subsequent wave phases, will be
presented. This will provide the information on not only spatial
but also temporal variations of the two critical coronal
parameters.

In the following section, we show major relevant results on
streamer waves of paper I, and present the radial profiles of the
Alfv\'en speed. The strength of the coronal magnetic field will be
deduced in the third section. Conclusions and discussion are given
in the fourth section of this paper, and the associated wave
dispersion analysis is presented in the Appendix.

\section{Phase speed observations and the deduced profiles of the Alfv\'en speed}
Observational features of the streamer wave event dated on 6 July
2004 were described in paper I. The wave profiles, as well as
heliocentric distances of five propagating wave troughs/crests,
were extracted from the running difference images of the white
light coronagraph observations. The corresponding phase speeds
were derived with a second-order polynomial fitting to the
relevant distance-time profiles. In Figure 1a, we re-show the
radial profiles of the first two sets of phase speeds,
corresponding to the outward propagation of the first observed
wave trough and the following crest which are referred to as P1
and P2 hereafter. These phase speeds will be used for further
seismological studies. From Figure 1a, we see that the phase speed
declines with increasing distance, e.g, for P1 (P2), the speed
decreases from 493 (428) km s$^{-1}$ at 3.3 (3.05) $R_\odot$ to
474 (411) at 5 $R_\odot$ and 415 (361) km s$^{-1}$ at 10
$R_\odot$. At a fixed distance, P1 moves faster than P2 by about
60 km s$^{-1}$. This difference of phase speeds was employed to
understand the increase of wavelength during the propagation of
the streamer wave.

The observed streamer wave was interpreted as the fast body kink
mode carried by the thin-sheet structure of the streamer stalk. To
deduce the mode phase speed in the plasma rest frame ($v_k$), one
has to determine the speed of the mean flow through which the mode
propagates. As mentioned previously, velocity measurements given
by about 80 blobs and shown in Figure 1b as shadows are used to
provide constraints on the speed of the solar wind along the
plasma sheet (Wang et al. 2000). The solid line inside the shadow
area represents the mean solar wind velocity. Subtracting the
shown solar wind velocities from the measured phase speeds in
Figure 1a, one gets $v_k$. To deduce the concerned Alfv\'en speed
from $v_k$, one still needs to determine their relationship. This
is done in the Appendix with a parameter study on the dispersion
relation of the fast kink body mode in a slab configuration of
magnetized plasmas under local approximations (Equation (11) of
Edwin {\&} Roberts 1982). From the study, we conclude that it is
appropriate to approximate the Alfv\'en speed in the region
surrounding the plasma sheet ($v_{Ae}$) by the equation
$v_{Ae}=v_k/\alpha$ where $\alpha$ is fixed to be 0.92.

The external Alfv\'en speed $v_{Ae}$ thus obtained are presented
in Figure 1c with light and dark shadow areas corresponding to
phases P1 and P2, respectively. The solid lines are associated
with the mean wind speed as plotted in Figure 1b. It can be seen
that $v_{Ae}$ decreases from 408 - 526 km s$^{-1}$ at 3.3
$R_\odot$ to 338 - 457 (175 - 294) km s$^{-1}$ at 5(10) $R_\odot$
for P1, and from 345 - 464 km s$^{-1}$ at 3.05 $R_\odot$ to 266 -
385 (116 - 235) km s$^{-1}$ at 5(10) $R_\odot$ for P2. Using the
same magnitude of the wind speed, the difference between the P1
and P2 $v_{Ae}$ is about 60 km s$^{-1}$. The timing difference
between the two phases is about half of the wave period, i.e.,
about half an hour. Therefore, it is suggested that the above
difference of $v_{Ae}$ is a result of the recovering process of
the CME-disturbed corona during this interval.

\section{Radial profiles of the magnetic field strength}
In the above section, we deduce the radial profiles of the
Alfv\'en speed in the region surrounding the plasma sheet. To
determine the corresponding magnetic field strength one needs to
specify the plasma density distribution in the region. This is
done through inversion of the LASCO pB data, as mentioned earlier.
Since the pB data were recorded at 21:00 on 6 July 2004 when the
streamer already got deflected and waved away from its
equilibrium, it is not possible to determine a sole direction of
the axis. We therefore make use of the pB data obtained one day
earlier for the purpose of inversion. This is equivalent to
assuming that the streamer density does not change appreciably
during the day from 21:00 5 July to 21:00 6 July.

Before showing the electron density profiles, we first discuss the
latitudinal variation of the pB data plotted in Figure 2b for 4
(solid) and 5 (dashed) $R_\odot$. From this figure, we have
determined the angular width of the plasma sheet to be about
3$^\circ$ (see the Appendix). It can be seen that outside of the
plasma sheet, the pB value, or mostly equivalently the electron
density at the same height, decreases gradually in a range of
about 10$^{\circ}$. This inhomogeneous density distribution
outside the plasma sheet is different from what was assumed in the
simplified slab model employed in the Appendix, where densities
distribute uniformly in both the interior and exterior of the
plasma sheet. To reconcile this discrepancy, we suggest that the
exterior parameters $n_e$, $v_{Ae}$, as well as $B_e$ to be
investigated, should be regarded as effective averages of the
inhomogeneously-distributed parameters outside the realistic
plasma sheet.

Keeping this in mind, we select two position angles (PA)
226$^{\circ}$ and 211$^{\circ}$, as already marked in Figures 2a
and 2b. The former PA lies near the streamer axis, the latter is
placed about 15 degrees away. The electron density profiles along
the PAs below 5 $R_\odot$ given by the pB inversion method and
those beyond determined by assuming the $r^{-2}$ dependence are
shown in Figure 2c with dashed (226$^\circ$) and dotted
(211$^\circ$) lines. Along PA $= 226^\circ$ the electron density
equals 7.19$\times10^5$ (1.67$\times10^5$) cm$^{-3}$ at 3 (5)
$R_\odot$, and along PA=211$^\circ$ the electron density is much
smaller being 7.82$\times10^4$ (1.56$\times10^4$) cm$^{-3}$ at the
same distance. The solid line represents the average of the two
densities, which are approximately half of the densities along the
streamer axis, being 3.99 $\times10^5$ (9.12$\times10^4$)
cm$^{-3}$ also at 3 (5) $R_\odot$. According to the above
discussion, we take the average density to be $n_e$, which is then
substituted into the obtained $v_{Ae}$ values to estimate the
strength of the magnetic field $B_e$. The limitations of this
approach will be discussed in the following section.

Radial profiles of $B_e$ corresponding to P1 and P2 are shown in
Figure 3 with light (upper) and dark (lower) shadow areas,
respectively. It can be seen that $B_e$ varies in between 0.096 -
0.123 G at 3.3 $R_\odot$, and decreases to 0.047 - 0.064
(0.012-0.020) G at 5(10) $R_\odot$ for P1, and decreases from
0.096 - 0.129 G at 3.05 $R_\odot$ to 0.037 - 0.053 (0.008-0.016) G
at 5(10) $R_\odot$ for P2. The variation of $B_e$ at a fixed
distance for a certain wave phase is a straightforward result of
the large range of the solar wind velocity used in the deduction
of $v_{Ae}$. A larger $B_e$ ($v_{Ae}$) corresponds to a slower
solar wind. Thus, the impact of the velocity of the solar wind
along the plasma sheet, which is not yet directly measurable, on
our diagnostic results can be directly read from the figures. The
solid lines in the middle of the shadow areas denote $B_e$
distributions associated with the mean velocity of the solar wind.
For comparison, we also plot the magnetic field strength profiles
decreasing according to the $r^{-2}$ dependence. We see that
deviations from radial expansion are not significant, so we
suggest that the deduced magnetic field expands basically
radially.

The magnetic field corresponding to P1 is generally stronger than
that corresponding to P2 if identical solar wind speeds are used
to deduce the field strength. The discrepancies of the two sets of
field strength can be regarded as the field temporal change over
the interval of half an hour. The change is suggested to be a
result of the recovering process of the CME-disturbed corona, as
proposed to explain the $v_{Ae}$ change in the above section. It
should be pointed out that for P1 and P2 we have used the same set
of wind parameters, while they do change temporally with the
recovering process. This contributes to the uncertainty of our
results. More discussions will be presented in the following
section.

In Figure 3, we also show other diagnoses on the magnetic field
strength in the corona with various symbols. The preliminary
results of $B_e$ of paper I are shown as stars, which are 0.045
(0.01) G at 5 (10) $R_\odot$ based on the P2 phase speed
measurement and the solar wind model of Chen {\&} Hu (2001). Other
results are mostly obtained employing radio methods. To be
specific, the strength-distance relationship in the heliocentric
range of 1.02 - 10 $R_\odot$ above active regions (dot-dashed
line) given by Dulk {\&} McLean (1978) are mainly based on radio
burst observations, the results of Vrsnak et al. (2002) and Cho et
al. (2007) presented as crosses and diamonds are deduced using the
band-splitting phenomenon of Type-II radio bursts, the results of
P\"{a}tzold et al. (1987) plotted as open circles with error bars
are given by the Faraday-rotation measurement of radio signals
emitted by the Helios spacecraft, similar results associated with
the extragalactic radio signals are included as triangles
(Spangler 2005) and squares (Ingleby et al. 2007) for heliocentric
distances of 6.2 $R_\odot$ (0.039 G) and 5 $R_\odot$ (0.046 -
0.052 G), respectively. The latest results obtained by Ramesh et
al. (2010) employing the low-frequency circularly polarized radio
emission at 1.5 (6 $\pm$ 2 G) and 1.7 (5 $\pm$ 1 G) $R_\odot$
inside a streamer structure are given as solid inverse triangles.
We note that the above list of diagnoses of the coronal field
strength is incomplete, and there exist many other estimates (see
e.g., references in Vrsnak et al. 2002). From Figure 3, it can be
seen that the magnitude and variation trend of our magnetic field
strength in the region surrounding the plasma sheet are basically
in line with others.

\section{Conclusions and discussion}
In this paper we provide a novel method to diagnose the Alfv\'en
speed and magnetic field strength in the corona based on the
observations of streamer waves, which propagate with phase speeds
consisting of two components. One is the speed of the wave mode in
the plasma rest frame, the other is the speed of the mean solar
wind. The method applies to heliocentric ranges from 3 to about 10
$R_\odot$ in the region surrounding the plasma sheet. To implement
the diagnosis, we first establish the connection between the
Alfv\'en speed and the observed phase speed, then we put
constraints on the solar wind velocities with blob measurements,
and determine the density distributions through the inversion of
the LASCO pB data. The obtained profiles of the magnetic field
strength are in line with other estimates in the corona.

Previous studies indicate that the magnetic flux tube experiences
a dramatic expansion in the neighborhood of the streamer cusp, and
a possible contraction beyond (e.g., Wang 1994; Bravo {\&} Stewart
1997; Chen {\&} Hu 2001, 2002; Hu et al. 2003; Li et al. 2006).
Till now, there exist no direct observational proofs of the above
peculiar feature of the magnetic field near the streamer cusp.
From the seismological study of this paper, we see that the
magnetic field along the plasma sheet expands more or less
radially starting from as near as 3 $R_\odot$. Therefore, the
mentioned intriguing expanding process of the flux tube occurs, if
it does, below this distance. This provides observational
constraints on relevant models.

According to our seismological studies on the basis of speed
measurements of the two wave phases, we find that the Alfv\'en
speed and magnetic field strength at a fixed distance decrease
with time. This has been suggested to be a result of the
recovering process of the CME-disturbed corona. In the process,
the magnetic field initially stretched out by the CME ejecta may
get relaxed through processes like magnetic reconnections, and the
evacuated corona may get refilled gradually through plasma heating
and resultant expansions. As mentioned in the above section, the
occurrence of this dynamic recovering process contributes directly
to the uncertainty of our results, since we have adopted identical
and steady solar wind parameters for the diagnoses associated with
the two phases. Due to the lack of direct measurements on these
parameters in the near-Sun region, it is currently not possible to
evaluate the impact of using time-dependent solar wind parameters
on our results.

Apart from the undeterminancy associated with solar wind densities
and velocities, there exist two other major factors contributing
to our diagnostic uncertainties. One is the error coming from the
phase speed measurements, which was already discussed and
estimated to be about $\pm10\%$ (or $\pm50$ km s$^{-1}$) in paper
I. This error will be passed directly to the evaluation of the
Alfv\'en speed and the magnetic field strength. Another factor
stems from the approximate relationship between the kink mode and
the external Alfv\'en speed. In the paper the relationship was
determined with the dispersion relation given by Edward {\&}
Roberts (1982) for a simplified magnetized plasma slab
configuration under Cartesian coordinates. This geometry is
different from the realistic spherical, inhomogeneous (both in the
radial and latitudinal directions), and time-dependent
streamer-plasma-sheet configuration. Therefore, future studies
should investigate the properties of the kink mode under more
realistic geometry. In addition, theoretical and numerical efforts
should continue to explore excitation conditions and propagation
properties of streamer waves in the process of CME-streamer
interaction, and determine the connection between the wave
properties and the background plasma properties. These works will
be of great benefit to future seismological studies with streamer
waves.

\acknowledgements The SOHO/LASCO data used here are produced by a
consortium of the Naval Research Laboratory (USA),
Max-Planck-Institut f\"{u}r Aeronomie (Germany), Laboratoire
d'Astronomie Spatiale (France), and the University of Birmingham
(UK). SOHO is a project of international cooperation between ESA
and NASA. We thank Dr. A. Vourlidas for helping us analyze the
LASCO pB data. This work was supported by grants NNSFC 40774094,
40825014, 40890162, A Special Fund for Public Welfare Industry
(meteorology): GYHY200806024, and A Foundation for the Author of
National Excellent Doctoral Dissertation of PR China (2007B24). B
Li is supported by the grant NNSFC 40904047, and LD Xia by
40774080 and 40974105.

\appendix

\section{Appendix: Dispersion diagrams of the fast kink body mode and the relationship of $v_k$ and $v_{Ae}$}

To determine the relationship between the phase speed of the fast
kink body mode and the Alfv\'en speed, we conduct a parameter
study on the dispersion relation in a slab configuration of
magnetized plasmas under local approximations (Equation (11) of
Edwin {\&} Roberts 1982), which is written as
$$
n_e(k^2v_{Ae}^2-\omega^2)m_0
\texttt{coth}(m_0x_0)+n_0(k^2v_{A0}^2-\omega^2)m_e=0
$$
where $\omega(k)$ is the wave frequency (number), $x_0$ is the
half width of the slab. The subscript ``0'' (``e'') represents
parameters for interior (exterior) of the plasma slab, $n$ being
the number density, the Alfv\'en speed $v_{A0,e}={B_{0,e}\over
\sqrt{\mu_0 m_p n_{0,e}}}$. The sound speed
$c_{s0,e}=\sqrt{{2\gamma k_B T_{0,e}\over m_p}}$ is implicitly
contained in the expressions of $m_0$ and $m_e$ whose definitions
are given in the above paper. In our study the polytropic index
$\gamma$ is taken to be 1.1 to calculate the sound speeds in the
heated coronal environment. For the fast body kink mode, it is
required that ${m_0}^2<0$ and $m_e>0$. To solve the dispersion
relation and to find out the relationship between the observed
phase speed and the Alfv\'en speed, it is necessary to specify all
coronal parameters required in the above dispersion relation,
including number densities, plasma temperatures, and magnetic
field strengths both inside and outside of the slab structure of
the plasma sheet. Different sets of coronal parameters at three
nominal heliocentric distances (3, 5, and 7 $R_\odot$) are
specified based on available observational diagnoses and
theoretical modelings.

The electron number density along a streamer axis was given by
many other authors using the inversion method of the LASCO-pB data
(e.g., Strachan et al. 2002, Uzzo et al., 2006). For the specific
streamer in question we take advantage of the pB inversion program
enclosed in the Solar Software system (SSW) to give the electron
density profile along the axis, which is taken to be the density
inside the slab $n_0$. The obtained density profile has been
presented in Figure 2. Here we repeat the density values which are
7.19$\times10^5$ (1.67$\times10^5$) cm$^{-3}$ at 3 (5) $R_\odot$.
The density at 7 $R_\odot$ is estimated to be 8.5$\times10^4$
cm$^{-3}$ by assuming an $r^{-2}$ dependence with $r$ representing
the heliocentric distance. The electron density outside the sheet
$n_e$ is given by half of the interior densities at the same
height for simplicity.

The temperature required to solve the dispersion relation is the
average of the proton and electron temperatures since the slab
model makes use of a single-fluid approximation. There exist
certain measurements on the kinetic temperatures (sum of the
thermal temperature and the contribution due to turbulent motions)
of protons along the streamer axis from about 1.75 $R_\odot$ to 5
$R_\odot$ (e.g., Strachan et al., 2002; Uzzo et al., 2006), which
indicate that the proton kinetic temperature decreases slowly with
radial distance from 1.58 MK at 3 $R_\odot$ to $1.24$ MK at 5
$R_\odot$. However, no electron temperature measurements are
available beyond 1.5 $R_\odot$ in the corona. Nevertheless, there
are numerical endeavors which manage to produce electron
temperature profiles in the solar wind along the streamer plasma
sheet (e.g., Chen et al., 2004; Li et al., 2006). These models
show that the electron temperature also decreases slowly from a
value larger than 1 MK at the coronal base to 4-6$\times$10$^5$ K
at about 10 $R_\odot$. Keeping these numbers in mind, we adopt two
sets of temperature profiles for the parameter study on the
dispersion relation, one set is for isothermal temperature
profiles (being 1 MK everywhere), the other set is for an
$r^{-1}$-dependent temperature profile with a prescribed value of
1.2 MK at 3 $R_\odot$.

Regarding the magnetic field strength, from the Faraday rotation
measurement of extragalactic radio sources Ingleby et al. (2007)
gave an estimate of about 0.05 G at 5 $R_\odot$ ``outside of the
region around the neutral line'', or in other words, outside the
streamer current-plasma sheet region. For our study, we take their
result to give one set of values of magnetic field strength
outside of the plasma sheet, i.e., $B_e$. Besides, we also use two
other field strengths at 5 $R_\odot$ different from their value by
$\pm25\%$, respectively. The field strengths at the rest of the
nominal distances are given using the $r^{-2}$ dependence. The
magnetic field strength inside the plasma sheet $B_0$ is
calculated from the total-pressure balance condition.

The yielded 18 dispersion curves are plotted all together in
Figure 4 with the solid-black (dotted-red, dashed-green) lines
corresponding to the nominal distance of 3 (5, 7) $R_\odot$. The
ordinate is the kink mode phase speed $v_k$ normalized by
$v_{Ae}$, the abscissa is the dimensionless wave number $kx_0$
with $k$ being the radial wave number and $x_0$ being the half
width of the plasma sheet, as already defined. It can be seen that
$v_k \approx v_{Ae}$ with a long wavelength or small $kx_0$, and
$v_k$ decreases monotonically with increasing $kx_0$. The values
of $k$ and $x_0$ for the concerned event will be specified in the
following paragraph.

In situ detections indicate that the angular width of the plasma
sheet structure is less than 3$^\circ$ (Borrini et al. 1981;
Goldstein et al. 1996), this is consistent with coronagraph
observations. In Figure 2a, we show the pB data obtained by LASCO
on 5 July 2004, one day prior to the streamer wave event. The two
arcs are at heliocentric distances of 4 and 5 $R_\odot$,  the pB
data along which are shown in Figure 2b. The data are normalized
by corresponding maximum pB values. We see that 3$^\circ$ is an
appropriate value for the angular width of the plasma sheet,
through which the width of the plasma sheet at a certain
heliocentric distance can be easily determined. The wave numbers
$k=2\pi/\lambda$ can be calculated given the wavelengths $\lambda$
being 2.0 (2.1, 2.4) $R_\odot$ at 3 (5, 7) $R_\odot$ according to
paper I. We then have $kx_0$ = 0.25, 0.39, and 0.48 for the three
distances. We see that the corresponding ratio of
$v_k/v_{Ae}(=\alpha)$ varies in a small range from 0.97 to 0.88,
apparently insensitive to the changing parameters.

From further calculations (not included in this paper), it is
clear that using other reasonable parameters in between, or lower
than the above two sets of temperatures, larger field strengths,
or smaller outside densities results in no observable effect on
the final value of $\alpha$. This is actually a result of a
property of the solved dispersion relation of the fast kink body
mode: as long as $v_{Ae}$ is larger than all other characteristic
speeds present in the dispersion relation, the phase speed of the
mode is smaller yet rather close to $v_{Ae}$, i.e., $\alpha$ is
smaller yet rather close to unity. This allows us to adopt a
uniform approximate relationship of $v_{Ae}$ and $v_k$ for
different coronal parameters: $ v_{Ae} = v_k/\alpha$, where
$\alpha$ is fixed to be a constant of 0.92 according to the above
parameter study. Once the value of $\alpha$ is decided, the
Alfv\'en speed in the region outside of the plasma sheet can be
easily deduced. However, it should be noted that the actual
morphology of the streamer plasma sheet is more complex than the
slab model used here. Therefore, the work presented is expected to
be improved by future wave analysis considering some more
realistic configuration.

\newpage
\begin{figure}
% \epsscale{1.}
 \includegraphics[width=0.8\textwidth]{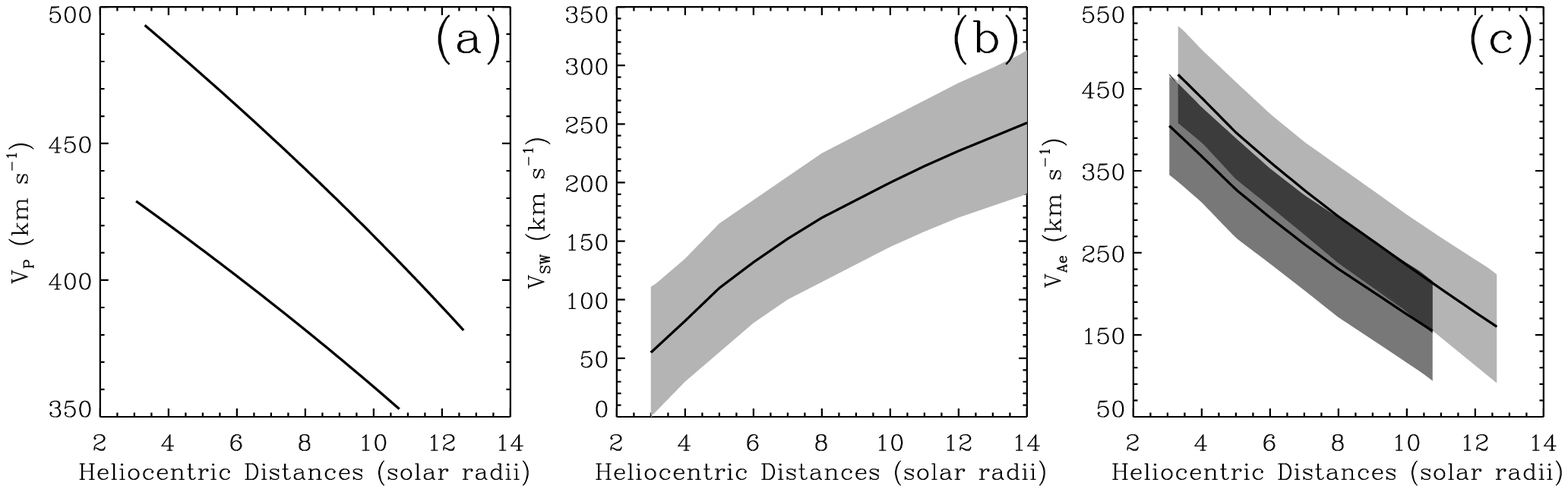}
\caption{(a) Radial profiles of propagation speeds of phases P1 and P2 of the streamer wave
observed on 6 July 2004 (taken from paper I); (b) speed variations
of the solar wind along the streamer plasma sheet, deduced from
the blob measurements (taken from Wang et al. 2000), the solid
line represents the mean speed; (c) light and dark areas represent
variation ranges of the exterior Alfv\'en speed $v_{Ae}$
corresponding to wave phases P1 and P2, the solid lines are
$v_{Ae}$ associated with the mean wind speed.\label{fig1}}
\end{figure}

\begin{figure}
% \epsscale{1.}
 \includegraphics[width=0.8\textwidth]{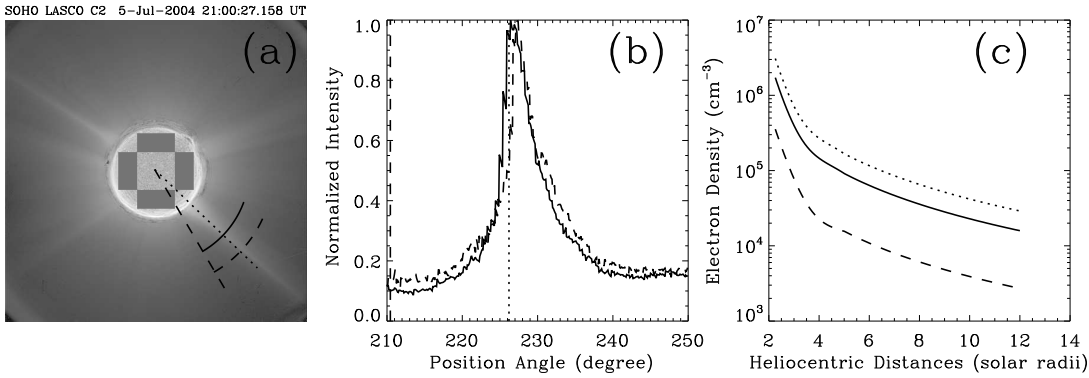}
\caption{(a) The pB distribution observed by LASCO, the two arcs are at heliocentric distances of 4
and 5 $R_\odot$, the dotted (dashed) line denotes a specific PA of
226$^\circ$ (211$^\circ$); (b) the solid (dashed) line presents
the latitudinal profiles of pB at 4 (5) $R_\odot$, the vertical
lines are the PAs shown in panel (a); (c) dotted and dashed lines
are the electron number densities given by the pB inversion
program for PA=226$^\circ$ and 211$^\circ$, respectively, the
solid line shows the average of them. Densities beyond 5 $R_\odot$
are determined according to an $r^{-2}$ dependence.\label{fig2}}
\end{figure}

\begin{figure}
% \epsscale{.8}
 \includegraphics[width=0.8\textwidth]{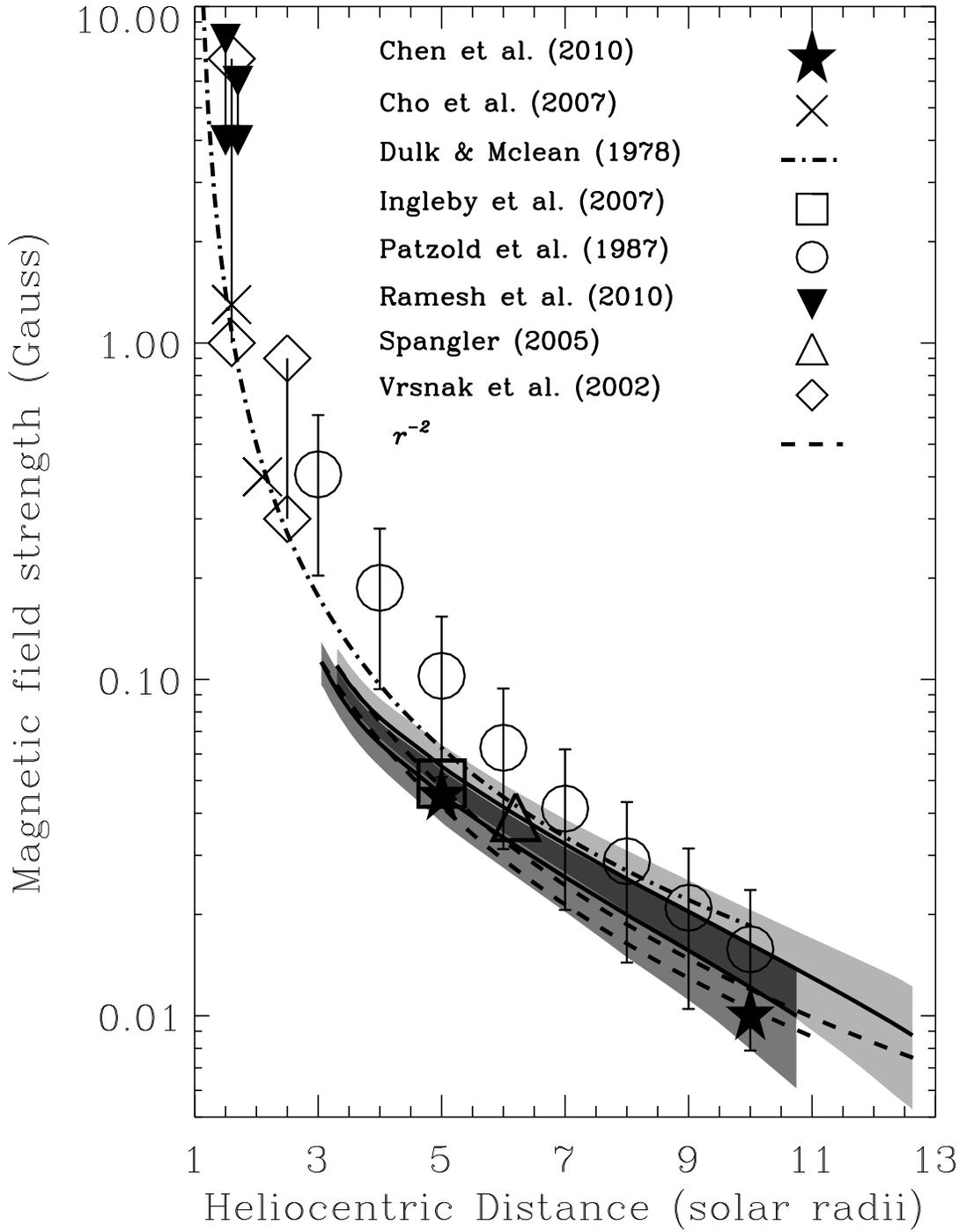}
\caption{Areas of shadow present the deduced radial profiles of the strength of the magnetic field
along the plasma sheet. The upper (lower) one is associated with
P1 (P2). The solid lines correspond to the mean solar wind speed
shown in Figure 1b, the dashed lines show the variation of the
field strength assuming the $r^{-2}$ dependence. Various symbols
represent other estimates on the coronal magnetic field strength.
\label{fig3}}
\end{figure}

\begin{figure}
% \epsscale{.8}
 \includegraphics[width=0.8\textwidth]{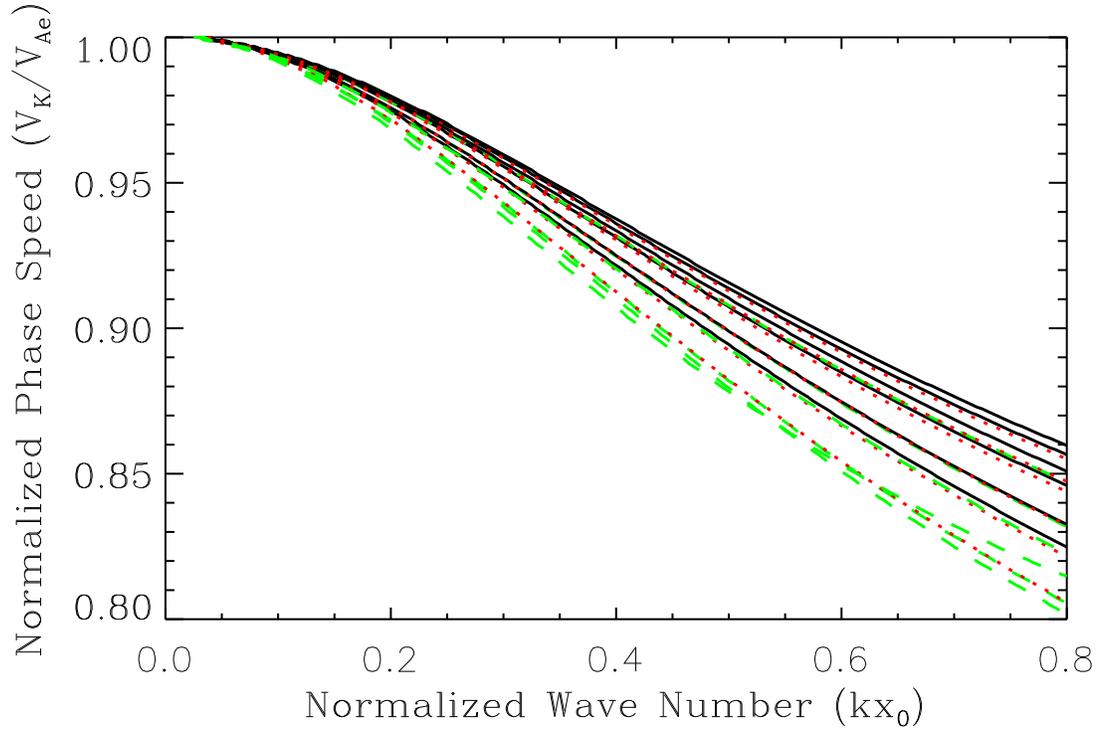}
\caption{18 dispersion curves of the fast kink body mode using various sets of prescribed coronal
parameters with the solid-black (dotted-red, dashed-green) lines
corresponding to parameters for the nominal distance of 3 (5, 7)
$R_\odot$, see the Appendix for details.\label{fig4}}
\end{figure}

\end{document}